\documentclass[twocolumn,showpacs,amsmath,amssymb]{revtex4}
\usepackage{appendix}
\usepackage{xcolor}
\usepackage{graphicx}
\usepackage{dcolumn}
\usepackage{bm}
\def\b{\begin{equation}}
\def\e{\end{equation}}
\begin{document}
\title{Screened plasmons of graphene near a perfect electric conductor}

\author{Afshin Moradi$^{1}$}
\email{a.moradi@kut.ac.ir}
\author{Nurhan T\"urker Tokan$^{2}$}
\email{nturker@yildiz.edu.tr}
\affiliation{%
\emph{$^{1}$Department of Engineering Physics, Kermanshah University of Technology, Kermanshah, Iran\\ $^{2}$Department of Electronic and Communications Engineering, Yildiz Technical University, Istanbul, Turkey  }
}%

\begin{abstract}
Screened plasmon properties of graphene near a perfect electric conductor are investigated using classical electrodynamics and a linearized hydrodynamic model that includes Fermi correction. A general expression for the dispersion relation of the mentioned screened plasmonic waves is given and illustrated graphically. The result indicates that for realistic wavenumbers, the dispersion relation of plasmonic waves of isolated graphene is almost unaffected by the Fermi correction, while this correction is an important factor for the screened plasmons of graphene near a perfect electric conductor, where it increases the frequency of surface waves. The results show that near the graphene neutrality point,
the surface wave has a linear dispersion with a universal speed close to $v_{\mathrm{F}}/\sqrt{2}$. Such linear dispersion for surface waves (also known as energy waves) appears to be a common occurrence when a splitting of plasma frequencies occurs, e.g. in the electron-hole plasma of graphene [W. Zhao \textit{et al}., Nature \textbf{614}, 688 (2023)]. Furthermore, analytical expressions for the energy parameters (the power flow, energy density, and energy velocity) of screened plasmons of the system are derived. Also, the analytical expressions are derived and analyzed for the damping function and surface plasmon and electromagnetic field strength functions of surface waves of the system with small intrinsic damping.

\end{abstract}

 \maketitle

\section{Introduction}
\label{sec:introduction}
Graphene is a two-dimensional (2D) material consisting of carbon atoms arranged in a hexagonal lattice, which was discovered by Novoselov \textit{et al}. \cite{K.S.N666} in 2004. Graphene has electrons that behave like massless Dirac fluid \cite{B.W318, E.H.H205418}, and therefore extraordinary properties can be observed in this 2D material. For example, graphene has carriers (i.e., electrons and holes) with extremely high mobility. Also, graphene supports the propagation of surface plasmon polariton (SPP) in the region from infrared to THz frequencies \cite{S.A.M016803, X.L351, P.A.D.G}. Furthermore, the low loss of SPPs of graphene up to mid-infrared frequencies also makes it a promising alternative for future applications \cite{X.L351}. 

The most important advantage of graphene in the plasmonics world is the tunability of surface plasmons because the density of carriers in graphene can be easily adjusted with doping and an electric gate. The conductivity characteristic of graphene \cite{M.M1052} and graphene's plasmonic properties \cite{A.N.G749} can be well explained by the hydrodynamic model derived by M\"{u}ller \textit{et al.} \cite{M.M025301} in the long-wavelength limit, i.e., $k\ll k_{\mathrm{F}}$, where $k_{\mathrm{F}}$ is the Fermi wavenumber in doped graphene and $k$ is the wavenumber of the plasmonic wave. Chaves \textit{et al.} \cite{A.J.C195438} investigated the excitation of
plasmonic waves of graphene in the presence of a fast-moving charge using the hydrodynamic model in the electrostatic approximation. Ferreira \textit{et al.} \cite{B.A.F033817} performed the quantization of graphene plasmons using the hydrodynamic model in the absence of losses for three graphene-based structures, i.e., a monolayer graphene, a bilayer graphene, and a graphene near a perfect electric conductor (PEC). The hydrodynamic Dirac fluid also shows interesting collective excitations, such as hydrodynamic bipolar plasmon polaritons that exhibit a coupled collective excitation of electromagnetic and electron-hole oscillations with the opposite motion of electrons and holes, and energy wave (also known as the demon mode), which is a quasi-acoustic mode in which the motion of relativistic electrons and holes are in the same direction \cite{D.S083715, A.L245153, Z.S3285, A.L053001, D.S121405, A.L115449, I.T144307, B.N.N167979, E.I.K245434, J.D023036, D.F941, B.N115402}. Importantly, more recently Zhao \textit{et al.} \cite{W.Z688} observed both hydrodynamic plasmons and the hydrodynamic energy waves of Dirac fluid based on new on-chip THz spectroscopy techniques, where the report by Zhao \textit{et al.} may reveal new opportunities to study the collective hydrodynamic excitations in graphene-based materials. 
\begin{figure}
\centering
  \includegraphics[width=8.2cm]{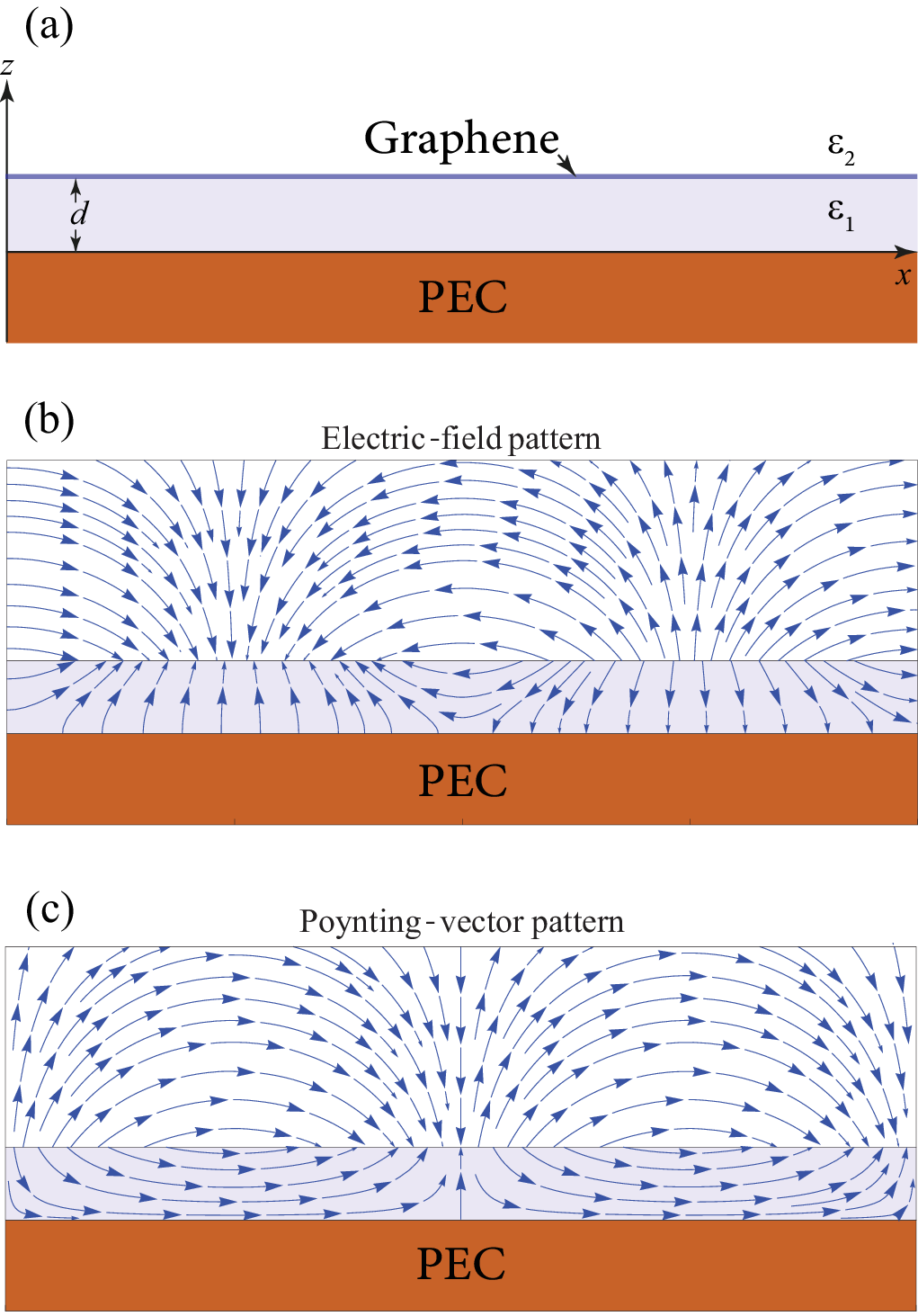}\\
  \caption{(a) Side view of the system under study. A monolayer graphene near a PEC. Monolayer graphene and PEC are separated from each other by a dielectric medium of thickness $d$ and dielectric constant $\varepsilon_{1}$, while the region
$z>d$ is a semi-infinite dielectric with the dielectric constant $\varepsilon_{2}$. (b) Snapshot of the electric-field pattern of screened $p$-polarized plasmon of graphene near a PEC in the $xz$ plane that shows the tangential component of the electric field vanishes at $z=0$. The magnitude of the electric field is oscillating in the $x$-direction but decreases away from the boundary. For the attenuation of the electric field, it can be found that the rate of hyperbolic sine decaying to zero is much faster than that of the exponential function. (c) Snapshot of the Poynting-vector pattern close to the boundary. The power flows in the $0<z<d$ region and in the $z>d$ region are in the same direction,
giving a net power flow to the right for a positive value of wavenumber $k$. One can see that $x$-component of the Poynting vector $S_{x}$ is maximum at the dielectric-PEC boundary and the Poynting vector is everywhere orthogonal to the electric field, as
expected on physical grounds.}\label{fig.1}
\end{figure}

One of the main advantages of the hydrodynamic model for the study of plasmonic waves of graphene is the possibility to include nonlocal and quantum effects in its plasmonic response without a high computational burden \cite{I.S.E133104}. Note that the condition $kc/(k_{\mathrm{F}}v_{\mathrm{F}})\gg 1$ must be satisfied for nonlocality to play an important role in the optical spectrum of graphene, where $v_{\mathrm{F}}$ is the Fermi speed in doped graphene, and $c$ is the speed of light in free space. Using the hydrodynamic model, we analyzed the characteristics of energy density and power flow of the $p$-polarized plasmonic waves of monolayer \cite{A.M63, A.M043103, A.M072114}, bilayer graphene \cite{A.M135}, and graphene on a conducting substrate \cite{A.M353}. However, to the best of our knowledge, no explicit calculation can be found for the energy behaviors of screened plasmons of graphene near a PEC. We note that the plasmonic properties of graphene near a PEC may present some new behaviors that make it appropriate for applications in the mid-infrared to a few THz range of frequencies. For instance, the results by Gu \textit{et al.} \cite{X.G071103} show that graphene near a PEC leads to additional field localization near graphene and may increase the amplification factor of plasmons, as shown by Morozov \textit{et al.} \cite{M.Y.M40}.

Therefore, in the present work, we wish to investigate the screened plasmon properties of graphene near a PEC for future applications. In this way, by using the linearized hydrodynamic model that includes Fermi
correction \cite{A.J.C195438} and classical electrodynamic formulations, we derive the general expressions for the dispersion relation, power flow, energy density, energy (group) velocity, damping function, and surface plasmon and electromagnetic field strength functions of the $p$-polarized surface waves of the system under investigation.

\section{Theory}
\label{2}

The side view of the system under study, i.e., a monolayer graphene near a PEC is shown in panel (a) of Fig. \ref{fig.1} in a Cartesian coordinate system with coordinates $(x,y,z)$. Note that the metal, typically gold or silver, may be modeled as a PEC for the frequency range of interest, i.e., from GHz to a few THz \cite{A.M2023, M.M.B494}. The monolayer graphene and PEC are separated by a dielectric of thickness $d$ and dielectric constant $\varepsilon_{1}$, whereas the region
$z>d$ is assumed to be a semi-infinite dielectric with the dielectric constant $\varepsilon_{2}$. The electronic behavior of the graphene layer is modeled as a 2D massless Dirac electron fluid and we assume that the equilibrium doping density of electrons in graphene is $n$. Now, let us consider the propagation of a plasmon polariton that is a $p$-polarized surface electromagnetic wave having $E_{x}$, $H_{y}$, and $E_{z}$ components along the $x$-direction at the boundary between two dielectric media. In this way, the homogeneous 2D massless Dirac electron fluid will be perturbed and can be regarded as a charged fluid with the
first-order perturbed values of the electron fluid density per unit area $n_{\mathrm{g}}(x,t)$ and electric current density flowing on the graphene surface $\textbf{J}(x,t)=j_{x}(x,t)\textbf{e}_{x}$, where $\textbf{e}_{x}$ being the unit vector along the $x$-axis. We assume that all physical quantities vary as $e^{i(k x-\omega
t)}$, where $k$ is the wavenumber (propagation constant) of the wave.

Based on the linear Drude  model with Fermi correction in the limit of large wavelengths, covering a range of frequencies from the mid-infrared to the THz, the electronic excitations on a doped graphene surface can be described by the following set of hydrodynamic equations \cite{A.J.C195438, A.S.P195437} \begin{figure}[!htb]
\centering
\includegraphics[scale=0.5]{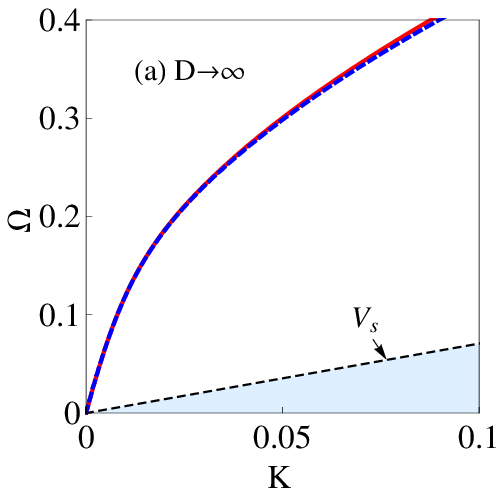}
\includegraphics[scale=0.5]{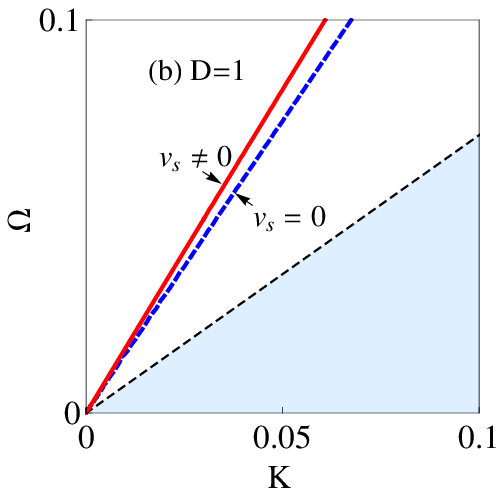}
\includegraphics[scale=0.5]{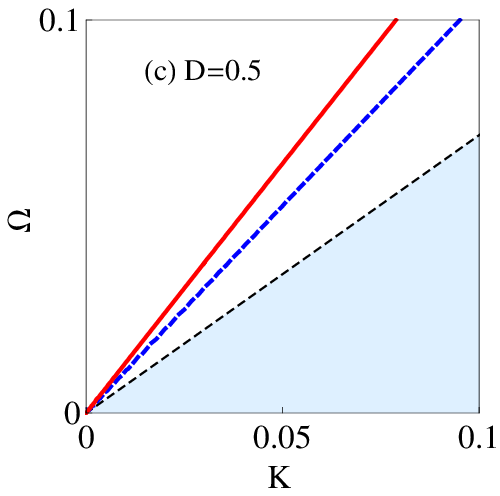}
\includegraphics[scale=0.5]{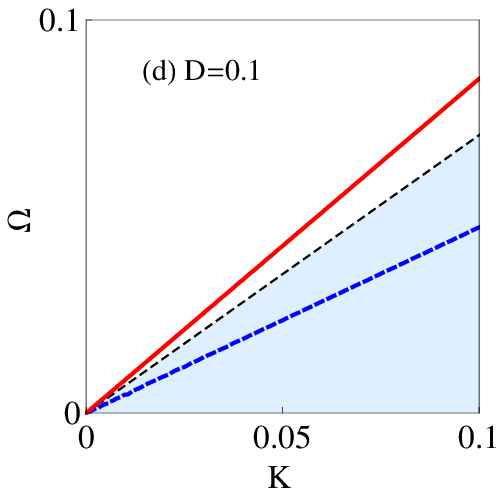}
\caption{ The dimensionless screened plasmon-polariton frequencies $\Omega=\omega/\omega_{0}$ of graphene near a PEC, as a function of $K=k/k_{0}$, for different values of the parameter $D=k_{0}d$, when $\varepsilon_{1}=\varepsilon_{\mathrm{SiO_{2}}}=3.9$, $\varepsilon_{2}=1$, and $n=n_{0}$ that means $k_{0}=k_{\mathrm{F}}$, and $\omega_{0}=\omega_{\mathrm{F}}$. The different
panels refer to (a) $d\rightarrow\infty$, (b) $d=1/k_{\mathrm{F}}$, (c) $d=0.5/k_{\mathrm{F}}$, and (d) $d=0.1/k_{\mathrm{F}}$. In each panel, the dashed blue line, and red line correspond to $s=0$ and $s\neq0$, respectively. The dashed black line marks a linear dispersion with
a velocity of $V_{\mathrm{s}}=v_{\mathrm{s}}/v_{\mathrm{F}}=1/\sqrt{2}$.}
\label{fig.2} 
\end{figure}
\begin{figure}[!htb]
\centering
\includegraphics[scale=0.5]{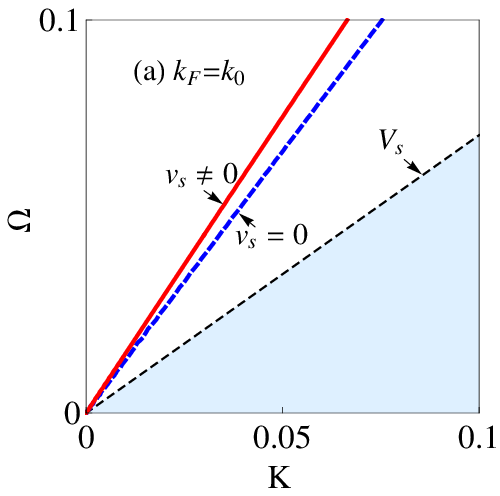}
\includegraphics[scale=0.5]{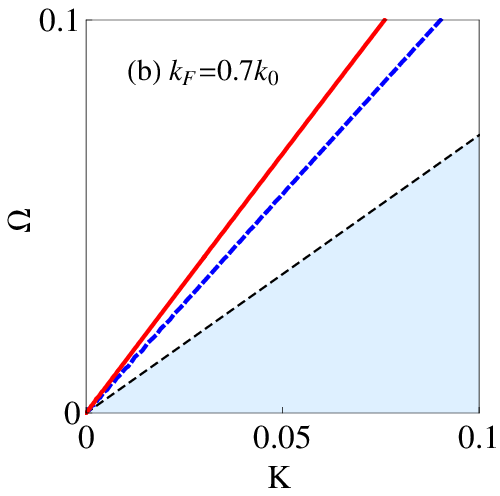}
\includegraphics[scale=0.5]{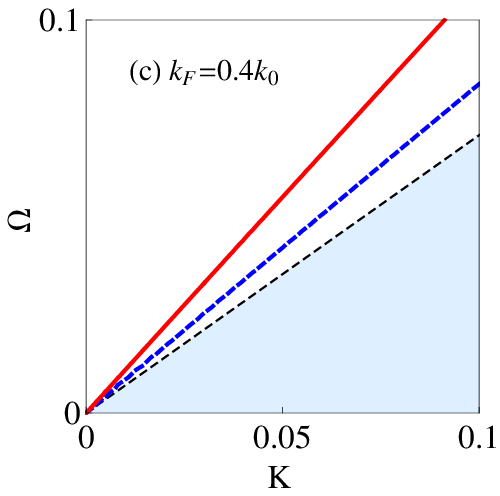}
\includegraphics[scale=0.5]{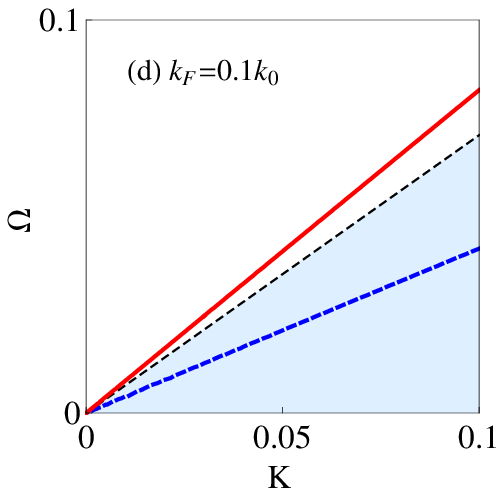}
\caption{ The dimensionless screened plasmon-polariton frequencies $\Omega=\omega/\omega_{0}$ of graphene near a PEC, as a function of $K=k/k_{0}$, for different values of the parameter $k_{\mathrm{F}}=\sqrt{\pi n}$, when $\varepsilon_{1}=\varepsilon_{\mathrm{SiO_{2}}}=3.9$, $\varepsilon_{2}=1$, and $d=0.8/k_{0}$. The different
panels refer to (a) $k_{\mathrm{F}}=k_{0}$, (b) $k_{\mathrm{F}}=0.7k_{0}$, (c) $k_{\mathrm{F}}=0.4k_{0}$, and (d) $k_{\mathrm{F}}=0.1k_{0}$. In each panel, the dashed blue line, and red line correspond to $s=0$ and $s\neq0$, respectively. The dashed black line marks a linear dispersion with
a velocity of $V_{\mathrm{s}}=v_{\mathrm{s}}/v_{\mathrm{F}}=1/\sqrt{2}$.}
\label{fig.3} 
\end{figure}
\b \label{1} -e\partial_{t}n_{\mathrm{g}}(x,t)+\partial_{x}j_{x}(x,t) =0\;,\e 
\b \label{2}\partial_{t}j_{x}(x,t)=\frac{D_{\mathrm{g}}}{\pi }E_{x}\big\vert_{z=d}+ev_{\mathrm{s}}^{2}\partial_{x}n_{\mathrm{g}}(x,t)\;,\e 
where $e$ is the electron charge, $D_{\mathrm{g}}=(e^{2}/\hbar^{2})E_{\mathrm{F}}$ is the Drude weight of graphene, $E_{\mathrm{F}}=\hbar \omega_{\mathrm{F}}$ (with $\omega_{\mathrm{F}}=v_{\mathrm{F}} k_{\mathrm{F}}$, and $\hbar=h/2\pi$, where $h$ is the Planck
constant) is the Fermi energy, $k_{\mathrm{F}}=\sqrt{\pi n}$ is the Fermi wavenumber and $v_{\mathrm{F}}\approx c/300$ is the Fermi speed in doped graphene, as mentioned before. In the right-hand side of Eq. \eqref{2}, the first term is the force on electrons due to the tangential component of the electric field, evaluated at the graphene surface $z=d$, and the second term shows the Fermi pressure in the 2D electron gas with
$v_{\mathrm{s}}=\upsilon_{\mathrm{F}}/\sqrt{2}$ that is the energy sound speed of Dirac electrons in graphene \cite{W.Z688}. Note that Eqs. \eqref{1} and \eqref{2} provides an
adequate description of the low-energy intraband electronic transitions in doped graphene
in the optical limit, specifically, for the wavenumbers $k\ll k_{\mathrm{F}}$.

Now, by eliminating the induced density $n$ from Eqs. \eqref{1} and \eqref{2}, and applying $j_{x}=\sigma_{\mathrm{g}} E_{x}$ (where $\sigma_{\mathrm{g}}$ is the conductivity of graphene), we find \cite{G.A.M1150} 
\b \label{3}\sigma_{\mathrm{g}}(k,\omega)=\dfrac{i}{\pi}\dfrac{\omega D_{\mathrm{g}}}{\omega^{2}-v_{\mathrm{s}}^{2} k^{2}}\;.\e

In order to determine the plasmonic properties of the system, we look for an evanescent $p$-polarized wave described
by an electric field of the form \cite{A.M}
$ \textbf{E}(x,z)=\left[ \textbf{e}_{x}E_{x}+\textbf{e}_{z}E_{z}\right] e^{ik x}$.
Note that we have considered $\partial \textbf{E}/\partial y=0 $, since the $x$- and $y$-directions are equivalent. The associated magnetic field has the form $ \textbf{H}(y)=\textbf{e}_{y} H_{y}e^{ik x}$, where we have omitted writing explicitly
a factor $\exp(-i\omega t)$ describing the time-dependence of the wave. Note that $H_{y}$ and $E_{z}$ can be determined if the non-zero longitudinal component $E_{x}$ is known. The Helmholtz equation for the x-component electric field $E_{x}$ of the $p$-polarized surface wave can be given by  
\b \label{4}\left[\dfrac{d^{2}}{d z^{2}}-\kappa_{\ell}^{2}\right] E_{x}(z)=0\;,   \e
 where 
$ \kappa_{\ell} =\left[k^{2}-\varepsilon_{\ell}k_{0}^{2}\right]^{1/2}$ with $\ell=1,2$, denotes the attenuation
constant in the regions $0<z<d$ and $z>d$, and $k_{0}=\omega/c$ is free space wavenumber. To solve Eq. \eqref{4}, we have to provide appropriate boundary conditions. With the electric conductivity of graphene, these boundary conditions at the surface $z=d$ can be written as
 \b \label{5}E_{x}\big\vert_{z=d+} =E_{x}\big\vert_{z=d-}\;, \e \b
 \label{6}H_{y}(z)\vert_{z=d+} -H_{y}(z)\vert_{z=d-}=-\sigma_{\mathrm{g}} E_{x}\big\vert_{z=d}\;, \e
which express the continuity and discontinuity of the tangential components of the
electric and magnetic fields, respectively, across the surface $z=d$. Also, the boundary condition satisfied by $E_{x}(z)$ at the surface $z=0$ is \b \label{7}E_{x}\big\vert_{z=0}=0\;,\e
that implies the tangential component of the electric field should vanish at $z=0$ as can be seen in panel (b) of Fig. \ref{fig.1}. With the above equations and also using $E_{z}=-(ik/\kappa_{\ell}^{2})\partial E_{x}/\partial z$, and $H_{y}=(i\omega\varepsilon_{0}\varepsilon_{\ell}/\kappa_{\ell}^{2})\partial E_{x}/\partial z$, we will investigate the screened plasmon-polariton properties of graphene near a PEC in the following sections.

\subsection{Dispersion relation}
\label{2.1}

We note that the surface field should decay for $z\rightarrow\infty$. Also, the presence of a PEC at $z=0$ implies
that $E_{x}(z=0)=0$. Therefore, the appropriate expressions for $E_{x}(z)$, are as follows: 
 \b \label{8}
E_{x}(z)=\left\{\begin{array}{clcr}
  A_{-} \sinh \kappa_{1}z \;, &
\mbox{$0\leq z\leq d$\;.}\
 \\ A_{+} e^{-\kappa_{2} z} \;, &
\mbox{$d\leq z$\;,}\  \\ \end{array}\right.\e
where the relations between the coefficients $A_{+}$ and $A_{-}$ can be determined from the
matching boundary conditions at the graphene surface, i.e., $z=d$. Use of Eqs. \eqref{4} and \eqref{8} in the boundary conditions \eqref{5} and \eqref{6} yields the condition that 
 \b \label{9} \dfrac{\varepsilon_{1}}{\kappa_{1}}\coth \kappa_{1}d+\dfrac{\varepsilon_{2}}{\kappa_{2}}+\dfrac{i\sigma_{\mathrm{g}} }{\omega\varepsilon_{0}}  =0 \;.\e
The roots of this transcendental equation, which can only
be solved numerically, provide the dispersion relation
of the screened plasmon-polaritons of graphene near a PEC. This dispersion relation has two tuning parameters: the graphene-PEC distance $d$, and the graphene sheet carrier density $n$, which controls the conductivity of graphene. When $d$ is very large such that $\coth \kappa_{1}d\approx 1$,
Eq. \eqref{9} reduces to the dispersion relation for the surface plasmon-polaritons supported by isolated monolayer graphene \cite{P.A.D.G, A.M}.

Let us note that the dispersion relation for the screened plasmon-polaritons of graphene near a PEC coincides with the quasi-acoustic plasmon-polaritons in symmetric bilayer graphene, provided $d=d_{\mathrm{bilayer}}/2$, where $d_{\mathrm{bilayer}}$ is the interlayer distance in the bilayer graphene. This fact can be understood in terms of image charges. That is why the screened plasmon-polariton introduced here is also called quasi-acoustic plasmon-polariton. Note that such a quasi-acoustic mode seems to be a common occurrence when a splitting of plasma frequencies happens due to the electrostatic interaction, e.g., in the electron-hole plasma of graphene \cite{D.S083715, W.Z688}, or in the coupling between the interlayer in the bilayer graphene \cite{A.M135}.

If we neglect the retardation effects, i.e., $k\gg k_{0}$, from Eq. \eqref{9} we find the dispersion relation
of the screened plasmons of graphene near a PEC, as
 \b \label{10}\omega=\left[  v_{\mathrm{s}}^{2} k^{2}+\dfrac{D_{\mathrm{g}}}{\pi\varepsilon_{0}}\dfrac{k }{\varepsilon_{1}\coth kd+\varepsilon_{2}}\right] ^{1/2} \;.\e
From Eq. \eqref{10}, we can distinguish two different
dimensionality regimes depending on two cases of $kd\gg 1$ and $kd\ll1$.
For $kd\gg 1$, where graphene and PEC decouple, we may use the asymptotic
expression $\coth kd\approx1$. Thus, the dispersion relation can be written as 
\b \label{11}\omega  =  \left[ v_{\mathrm{s}}^{2} k^{2}+\dfrac{D_{\mathrm{g}}}{\pi\varepsilon_{0}\left(\varepsilon_{1}+\varepsilon_{2} \right) }k\right]  ^{1/2} \approx \sqrt{\dfrac{D_{\mathrm{g}}}{\pi\varepsilon_{0}\left(\varepsilon_{1}+\varepsilon_{2} \right) }k}  \;.\e
which is exactly the same as the well-known dispersion relation of the surface plasmons of isolated graphene in free space, when $\varepsilon_{1}=1=\varepsilon_{2}$ \cite{A.J.C195438}, with the approximate result valid for realistic ($k\ll k_{\mathrm{F}}$ ) wavenumbers. This result means that the dispersion relation of an isolated graphene is almost unaffected by the Fermi correction, i.e., the internal pressure force of the electron [the term with $v_{\mathrm{s}}^{2}$ in Eq. \eqref{3}].

On the other hand, for $kd\ll 1$ we may use the asymptotic
expression $\coth kd\approx 1/kd$. Thus, the dispersion relation can be written as 
 \b \label{12}\omega=\left[ v_{\mathrm{s}}^{2} +\dfrac{D_{\mathrm{g}}}{\pi\varepsilon_{0}}\dfrac{d }{\varepsilon_{1}+\varepsilon_{2}kd}\right] ^{1/2}k\approx\left[  v_{\mathrm{s}}^{2} +\dfrac{ D_{\mathrm{g}}d}{\pi\varepsilon_{0}\varepsilon_{1}}\right]^{1/2}k \;.\e
As a new interesting result, it is clear that in this case, the internal pressure force of the electron is an important term and increases the frequency of surface plasmons. In fact, for
graphene near a PEC, the dispersion is strongly dependent on $d$. We note that Chaves \textit{et al.} \cite{A.J.C195438} showed that for $d$ about $1.5$nm, the screened plasmons of graphene near a PEC can appear in the mid-infrared with a wavenumber of the order of $200\mu$m$^{-1}$ (corresponding to a $\lambda_{\mathrm{spp}}=2\pi/k\approx30$nm). For a Fermi energy of graphene about $E_{\mathrm{F}}=0.4$eV, we find $kc/(k_{\mathrm{F}}v_{\mathrm{F}})\sim100$, which places graphene in the strong nonlocal regime. 

Also, the phase and group velocities of screened plasmons of the system can be obtained from Eq. \eqref{10}. For the phase velocity, we have
\b \label{13}v_{\mathrm{phase}}=\left[  v_{\mathrm{s}}^{2} +\dfrac{D_{\mathrm{g}}}{\pi\varepsilon_{0}k}\dfrac{1 }{\varepsilon_{1}\coth kd+\varepsilon_{2}}\right] ^{1/2} \;,\e
while for the group velocity by derivation of Eq. \eqref{10} with respect to $\omega$, we find
\b \label{14}v_{\mathrm{group}}=\dfrac{ v_{\mathrm{s}}^{2} k+\dfrac{D_{\mathrm{g}}}{2\pi\varepsilon_{0}}\dfrac{\varepsilon_{1}\coth kd+\varepsilon_{2}+\varepsilon_{1}\dfrac{kd}{\sinh^{2} kd }}{\left[ \varepsilon_{1}\coth kd+\varepsilon_{2}\right]^{2} }}{\left[  v_{\mathrm{s}}^{2} k^{2}+\dfrac{D_{\mathrm{g}}}{\pi\varepsilon_{0}}\dfrac{k }{\varepsilon_{1}\coth kd+\varepsilon_{2}}\right] ^{1/2}} \;.\e
To see clearly the character of the dispersion relation for the screened plasmon-polariton of graphene near a PEC, first let us introduce the dimensionless variables $K=k/k_{0},\Omega=\omega/\omega_{0}$, $V_{\mathrm{s}}=v_{\mathrm{s}}/v_{\mathrm{F}}=1/\sqrt{2}$, $V_{\mathrm{c}}=c/v_{\mathrm{F}}\approx300$, $D=k_{0}d$, where $k_{0}=\sqrt{\pi n_{0}}$, and $\omega_{0}=v_{\mathrm{F}} k_{0}$. Note that for $n_{0}=n$, we have $k_{0}=k_{\mathrm{F}}$ and $\omega_{0}=\omega_{\mathrm{F}}$. Now, in Fig. \ref{fig.2}, we show the dependence of the dimensionless frequency $\Omega=\omega/\omega_{0}$ on the dimensionless variable $K=k/k_{0}$, for different values of the parameter $D=k_{0}d$, when $\varepsilon_{1}=\varepsilon_{\mathrm{SiO_{2}}}=3.9$, $\varepsilon_{2}=1$, and $n=n_{0}$ that means $k_{0}=k_{\mathrm{F}}$, and $\omega_{0}=\omega_{\mathrm{F}}$. One can see that the behavior
of the screened plasmon-polariton depends on the value of $d$, where the decreasing thickness of the spacer layer $d$, red-shifts the frequency of the surface wave. More importantly, it can be seen that with the decreasing thickness of the spacer layer $d$, internal interaction force plays an important role in the dispersion relation of the surface wave for $k\ll k_{\mathrm{F}}$. We observe that in the presence of the nonlocal effects, screened plasmon-polariton of the system has a phase velocity that can be made arbitrarily close to the energy sound speed of Dirac
electrons in graphene by tuning the spacer layer $d$. Furthermore, from Fig. \ref{fig.2}, we observe for the constant values of frequency and graphene sheet carrier density, the wavelength of screened plasmon-polariton considerably
decreases when the PEC draws near to graphene, as can be easily concluded from Eq. \eqref{12}. 

The effect of the graphene sheet carrier density $n$ on the dispersion relation for the screened plasmon-polariton of graphene near a PEC is shown in Fig. \ref{fig.3}. It can be seen that as the carrier density decreases to lower values, the surface wave frequency decreases. It is clear that at extremely low charge density, the
surface wave has a velocity close to $v_{\mathrm{F}}/\sqrt{2}$. This means that near the graphene neutrality point,
the surface wave has a linear dispersion with a universal speed close to $v_{\mathrm{F}}/\sqrt{2}$ \cite{W.Z688}. One may conclude from panel (d) of Fig. \ref{fig.3} that near the neutral point of graphene, the local model shows an incorrect result. Also, from Fig. \ref{fig.3}, it is clear that for the constant values of $\omega$ and $d$, the wavelength of screened plasmon-polariton considerably
decreases with a decrease in the graphene sheet carrier density, as can be easily seen from Eq. \eqref{12}.

\subsection{Power flow} 
\label{2.2}
For the power flow density associated with a surface wave of graphene near a PEC, we have, in the three media,
\b \label{15}
\textbf{S}=\left\{\begin{array}{clcr}
\textbf{E}_{-}\times\textbf{H}_{-}\;, &
\mbox{$0<z<d$\;,}\
   \\ \textbf{E}_{+}\times\textbf{H}_{+}\;, &
\mbox{$d<z$\;,}\
 \\ \end{array}\right.\e
where subscripts $-$ and $+$ denote the regions below and above the graphene layer, and for $z=0$ using Eq. \eqref{A4} in Appendix, we have $ S_{\mathrm{g}x}=-\frac{\pi e}{D_{\mathrm{g}}} v_{\mathrm{s}}^{2} n_{\mathrm{g}}j_{x}$. After the elimination of $j_{x}=\sigma_{\mathrm{g}} E_{x}$ and $n_{\mathrm{g}}=-(k/e\omega)j_{x}$, and also using $A_{+}=A_{-}\sinh (\kappa_{1}d)\exp(\kappa_{2}d)$, the cycle-averaged $x$-components of $\textbf{S}$ in Eq. \eqref{15}, and also on the surface of graphene can be written as 
\b \label{16}
S_{x}=\dfrac{\varepsilon_{0}k\omega}{2}A_{-}^{2}\left\{\begin{array}{clcr}
\dfrac{\varepsilon_{1}}{\kappa_{1}^{2}}\cosh^{2} \kappa_{1}z\;, &
\mbox{$0<z<d$\;,}\  \\  \dfrac{\varepsilon_{2}}{\kappa_{2}^{2}}\sinh^{2} \kappa_{1}d\;e^{-2\kappa_{2}(z-d)} \;, &
\mbox{$d<z$\;,}\
\\ \end{array}\right.\e
\b \label{16f} S_{\mathrm{g}x}=\dfrac{\varepsilon_{0}k\omega}{2}A_{-}^{2} \frac{\pi }{\varepsilon_{0}D_{\mathrm{g}}\omega^{2}} v_{\mathrm{s}}^{2} \vert\sigma_{\mathrm{g}}\vert^{2}  \sinh^{2} \kappa_{1}d\;.\ \e
The total power flow density associated with the $p$-polarized surface wave is determined by integration over $z$. The power flow through an area in the $yz$ plane of infinite length in the $z$-direction and unit width in the $y$-direction is
\begin{multline} \label{17}
\left\langle S_{x}\right\rangle =\dfrac{\varepsilon_{0}k\omega}{4}\sinh^{2} \kappa_{1}d\;A_{-}^{2}\left[\dfrac{\varepsilon_{1}}{\kappa_{1}^{2}}\dfrac{d}{\sinh^{2} \kappa_{1}d}\right. \\ \qquad \left.+\dfrac{\varepsilon_{1}}{\kappa_{1}^{3}}\coth \kappa_{1}d +\dfrac{\varepsilon_{2}}{\kappa_{2}^{3}}+\dfrac{2\pi }{\varepsilon_{0}D_{\mathrm{g}}\omega^{2}} v_{\mathrm{s}}^{2} \vert\sigma_{\mathrm{g}}\vert^{2}  \right]  \;, \end{multline}
where $\left\langle \cdots\right\rangle\equiv\int_{0}^{+\infty}\cdots dz$. This total power flow density (per unit width) is positive, as can be seen in panel (c) of Fig. \ref{fig.1}.

\subsection{Energy distribution}
\label{2.3}
For the cycle-averaged energy density distribution associated with a surface wave of graphene near a PEC, we have,  \b \label{18}
U=\dfrac{1}{4}\left\{\begin{array}{clcr}
\varepsilon_{0} \varepsilon_{1} \vert\textbf{E}_{-}\vert^{2}+\mu_{0}\vert\textbf{H}_{-}\vert^{2}\;, &
\mbox{$0<z<d$\;,}\ 
 \\ \varepsilon_{0}\varepsilon_{2}\vert\textbf{E}_{+}\vert^{2}+\mu_{0}\vert\textbf{H}_{+}\vert^{2}\;, &
\mbox{$d<z$\;,}\  \\ \end{array}\right.\e
where for $z=0$ using Eq. \eqref{A3} in Appendix, we have \b \label{18f} U_{\mathrm{g}} =\frac{1}{4}\frac{\pi}{D_{\mathrm{g}}}\left(\vert j_{x}\vert^{2}+e^{2} v_{\mathrm{s}}^{2}\vert n_{\mathrm{g}}\vert^{2} \right)\;.\e Then Eqs. \eqref{18} and \eqref{18f} yield,
\b \label{19}
U=\dfrac{\varepsilon_{0}}{4}A_{-}^{2}\left\{\begin{array}{clcr}
 \varepsilon_{1} \left[ \sinh^{2} \kappa_{1}z+\dfrac{k^{2} +\varepsilon_{1}\frac{\omega^{2}}{c^{2}}}{\kappa_{1}^{2}}\cosh^{2} \kappa_{1}z \right] \;, &
\mbox{$z<d$\;,}\ 
\\ 2 \varepsilon_{2} \dfrac{k^{2}}{\kappa_{2}^{2}}\sinh^{2} \kappa_{1}d\; e^{-2\kappa_{2}(z-d)}\;, &
\mbox{$z>d$\;,}\ \\ \end{array}\right.\e
\b \label{19f} U_{\mathrm{g}}=\dfrac{1}{4}A_{-}^{2} \frac{\pi}{D_{\mathrm{g}}} \frac{\omega^{2}+v_{\mathrm{s}}^{2}k^{2}}{\omega^{2}}\vert\sigma_{\mathrm{g}}\vert^{2}  \sinh^{2} \kappa_{1}d\;,\e
where all contributions to the energy density are positive. The total energy density associated with the  screened plasmon-polaritons of graphene near a PEC is again determined by integration over $z$, the energy per unit surface area is \begin{figure}[!htb] 
\centering
\includegraphics[scale=0.47]{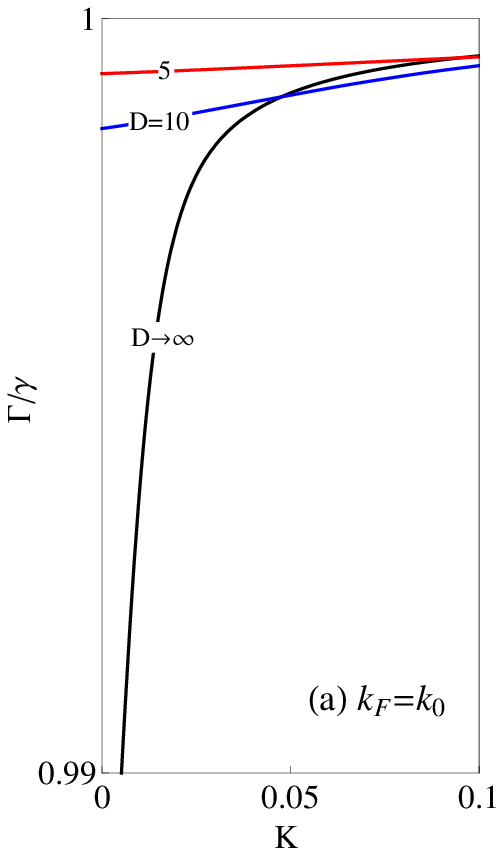}
\includegraphics[scale=0.47]{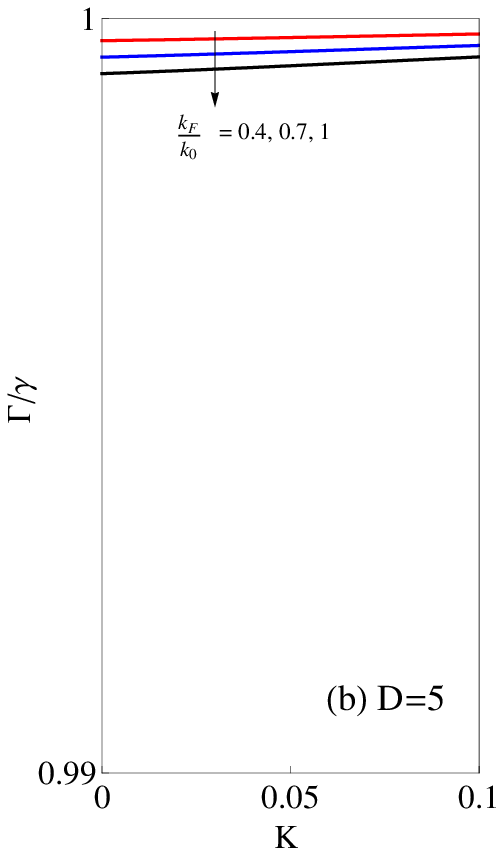}
\caption{ The wavenumber dependence of dimensionless relaxation rate of the long-wavelength screened plasmon-polaritons of graphene near a PEC as a function of $K=k/k_{0}$, when $\varepsilon_{1}=\varepsilon_{\mathrm{SiO_{2}}}=3.9$, $\varepsilon_{2}=1$. (a) For different values of the parameter $D=k_{0}d$, when $n=n_{0}$. (b) For different values of the parameter $k_{\mathrm{F}}=\sqrt{\pi n}$, when $D=5$. }
\label{fig.4} 
\end{figure}
\begin{figure}[!htb] 
\centering
\includegraphics[scale=0.47]{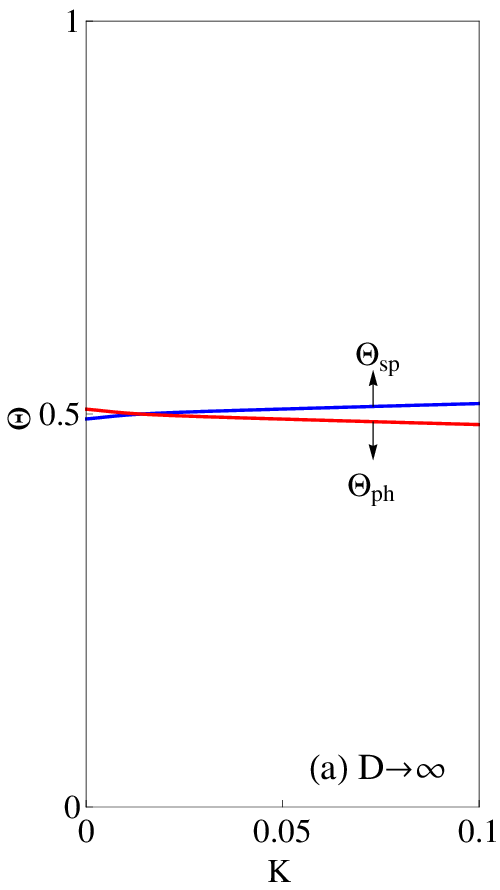}
\includegraphics[scale=0.47]{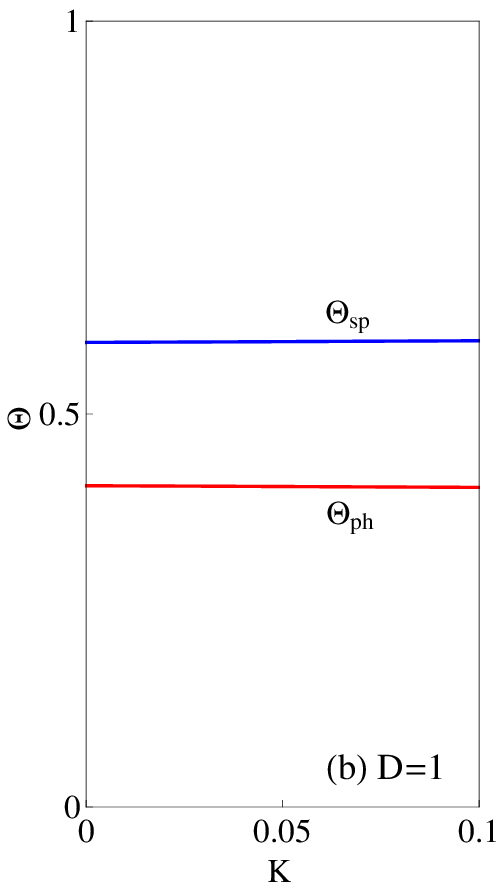}
\caption{The wavenumber dependence of the strength functions $\Theta_{\mathrm{sp}}$ and $\Theta_{\mathrm{ph}}$ of the screened plasmon-polaritons of graphene near a PEC, as a function of $K=k/k_{0}$, when $\varepsilon_{1}=\varepsilon_{\mathrm{SiO_{2}}}=3.9$, $\varepsilon_{2}=1$ for different values of the parameter $D=k_{0}d$, when $n=n_{0}$. }
\label{fig.5} 
\end{figure}
\begin{multline} \label{20}
\left\langle U\right\rangle =\dfrac{\varepsilon_{0}}{4}A_{-}^{2}\sinh^{2} \kappa_{1}d\left[ \frac{\varepsilon_{1}^{2}}{\kappa_{1}^{2}}\dfrac{\omega^{2}}{c^{2}}  \dfrac{d}{\sinh^{2} \kappa_{1}d}+k^{2}\dfrac{\varepsilon_{1} }{\kappa_{1}^{3}}\coth \kappa_{1}d \right. \\ \qquad \left.+ k^{2} \dfrac{\varepsilon_{2}}{\kappa_{2}^{3}}+\frac{\pi}{\varepsilon_{0}D_{\mathrm{g}}} \frac{\omega^{2}+v_{\mathrm{s}}^{2}k^{2}}{\omega^{2}}\vert\sigma_{\mathrm{g}}\vert^{2}  \right] \;.\end{multline}
Let us note that in a recent work, Morozov and Popov \cite{M.Y.M22209} prepared a concept of a terahertz waveguide plasmon amplifier based on a metal groove with active graphene. 
In this way, they used concepts of plasmon energy to gain insight into the physical origins of terahertz waveguide plasmon amplification. However, in their results, we cannot see the contribution of the energy density on the graphene surface, as shown by Eq. \eqref{19f}. Fortunately, there is an easy way to check such results. In fact, we should find the group and energy velocities of the wave under consideration, when losses are neglected. If the group velocity is the same as the energy velocity, then the obtained energy formulas are correct, as we are going to check this important issue for our results in the following subsection.

\subsection{Energy velocity}
\label{sec:2.4}
The energy velocity of a surface wave of graphene near a PEC is given as the ratio of the total power flow density (per unit width) and total energy density (per unit area), such as 
\begin{multline} \label{21} v_{\mathrm{energy}}=\omega k\\\dfrac{\frac{\varepsilon_{1}}{\kappa_{1}^{2}}\frac{d}{\sinh^{2} \kappa_{1}d}+\frac{\varepsilon_{1}}{\kappa_{1}^{3}}\coth \kappa_{1}d +\frac{\varepsilon_{2}}{\kappa_{2}^{3}}+\frac{2\pi }{\varepsilon_{0}D_{\mathrm{g}}\omega^{2}} v_{\mathrm{s}}^{2} \vert\sigma_{\mathrm{g}}\vert^{2}}{ \frac{\varepsilon_{1}^{2}}{\kappa_{1}^{2}}\frac{\omega^{2}}{c^{2}}\frac{d}{\sinh^{2} \kappa_{1}d} +k^{2}\frac{\varepsilon_{1} }{\kappa_{1}^{3}}\coth \kappa_{1}d +  k^{2}\frac{\varepsilon_{2}}{\kappa_{2}^{3}}+\frac{\pi}{\varepsilon_{0}D_{\mathrm{g}}} \frac{\omega^{2}+v_{\mathrm{s}}^{2}k^{2}}{\omega^{2}}\vert\sigma_{\mathrm{g}}\vert^{2}}\;.\end{multline}
If we neglect the retardation effects, i.e., $k\gg k_{0}$, from Eq. \eqref{22} we find
\b \label{22}v_{\mathrm{energy}}=\dfrac{ v_{\mathrm{s}}^{2} k+\dfrac{D_{\mathrm{g}}}{2\pi\varepsilon_{0}}\dfrac{\varepsilon_{1}\coth kd+\varepsilon_{2}+\varepsilon_{1}\dfrac{kd}{\sinh^{2} kd }}{\left[ \varepsilon_{1}\coth kd+\varepsilon_{2}\right]^{2} }}{\left[  v_{\mathrm{s}}^{2} k^{2}+\dfrac{D_{\mathrm{g}}}{\pi\varepsilon_{0}}\dfrac{k }{\varepsilon_{1}\coth kd+\varepsilon_{2}}\right] ^{1/2}} \;,\e
which is identical to the group velocity obtained in Eq. \eqref{14}.
This equality confirms the correctness of the presented results. In fact, in general, the group velocity is equal to the energy velocity in the absence of damping \cite{A.M18373, A.M143901, A.M10760}.

\subsection{Damping property}
\label{sec:2.5} 
Now, we study the damping function of surface waves of the system. To obtain an analytical expression for the damping function of the surface waves of graphene near a PEC we use the perturbative method proposed by Loudon \cite{R.L233} and Nkoma \textit{et al}. \cite{J.N3547}. Such a procedure enables us to calculate the true surface wave damping rate to the first order in the damping parameter $\gamma$, introduced to describe the intrinsic damping of crystal oscillations. Also, this theory enables us to discuss both the propagation length and the lifetime of a surface wave. The advantage of the perturbative method is that the damping properties result from the calculation of real dispersion relations. The plasmonic damping parameter or relaxation rate $\Gamma(k,\omega)$ of the present case may be determined by the following procedure. The kinetic and total energy densities (per unit area) $U_{\mathrm{gk}}$, and $\left\langle U\right\rangle$ are first calculated in the absence of damping. If a small amount of damping is now reintroduced, the surface energy relaxation rate to the lowest order in $\gamma$ is
$\Gamma(k,\omega)=2\gamma U_{\mathrm{gk}}/\left\langle U\right\rangle$.
where from Eq. \eqref{A3} we have
$ U_{\mathrm{gk}}=\dfrac{\varepsilon_{0}}{4}A_{-}^{2}\frac{\pi}{\varepsilon_{0}D_{\mathrm{g}}} \vert\sigma_{\mathrm{g}}\vert^{2}  \sinh^{2} \kappa_{1}d$.
Therefore, we get 
\begin{multline} \label{23} \Gamma(k,\omega)\\=\dfrac{2\gamma\frac{\pi}{\varepsilon_{0}D_{\mathrm{g}}} \vert\sigma_{\mathrm{g}}\vert^{2}  }{ \dfrac{d \frac{\varepsilon_{1}^{2}}{\kappa_{1}^{2}}\frac{\omega^{2}}{c^{2}} }{\sinh^{2} \kappa_{1}d}+k^{2}\frac{\varepsilon_{1} }{\kappa_{1}^{3}}\coth \kappa_{1}d + k^{2} \frac{\varepsilon_{2}}{\kappa_{2}^{3}}+\frac{\pi \vert\sigma_{\mathrm{g}}\vert^{2}}{\varepsilon_{0}D_{\mathrm{g}}} \frac{\omega^{2}+v_{\mathrm{s}}^{2}k^{2}}{\omega^{2}}  }\;. \end{multline}
Let us note that the frequency and wavenumber dependence of the damping function comes from the retarded part of the plasmonic waves and it is easy to find that, in the nonretarded limit, the total energy density (per unit area) becomes twice as large as the kinetic energy density (per unit area) of the system. As a consequence, the damping function of surface
waves of the system equals $\gamma$, i.e., it becomes a constant. Also, let us note that the surface wave lifetime $T$ is simply the inverse of Eq. \eqref{23}, i.e., $T(k,\omega)=\Gamma^{-1}(k,\omega)$, while their propagation length is given by $ L(k,\omega)=v_{\mathrm{group}}\Gamma^{-1}(k,\omega)$, where $v_{\mathrm{group}}$ can be found from Eq. \eqref{21}. 

By using \eqref{23}, the damping rate of the long-wavelength screened plasmon-polaritons of graphene near a PEC, in terms of the dimensionless variables are presented in Fig. \ref{fig.4} when $\varepsilon_{1}=\varepsilon_{\mathrm{SiO_{2}}}=3.9$, $\varepsilon_{2}=1$. It is clear that the damping function of long-wavelength screened plasmon-polaritons is approximately equal to $\gamma$ for a large value of $k$. From panel (a) one can see that by decreasing values of $D$ for a fixed value of $k_{\mathrm{F}}$, the screened plasmon-polaritons relaxation rate increases sharply for a low value of $k$. On the other hand, from panel (b) it is obvious that for a fixed value of $D$, by decreasing values of $k_{\mathrm{F}}$, the screened plasmon-polaritons relaxation rate increases.

\subsection{Surface plasmon and electromagnetic field strength functions}

\label{sec:2.6}
Since a screened plasmon-polariton is a coupled optical plasmon-photon wave, we can introduce strength
functions that characterize the quantitative compositions of the mixed wave. We
have $\left\langle U_{\mathrm{sp}}\right\rangle/\left\langle U\right\rangle+\left\langle U_{\mathrm{ph}}\right\rangle/\left\langle U\right\rangle=1$, where $\left\langle U_{\mathrm{ph}}\right\rangle$ is the total photon energy density and $\left\langle U_{\mathrm{sp}}\right\rangle$ is the
total surface plasmon energy density associated with the screened
plasmon-polariton. Note that $\left\langle U\right\rangle$ is the sum of the integrated energy densities, i.e., Eq. \eqref{20}, and also
\begin{multline} \label{24}
\left\langle U_{\mathrm{ph}}\right\rangle =\dfrac{\varepsilon_{0}}{4}A_{-}^{2}\sinh^{2} \kappa_{1}d\left[ \frac{\varepsilon_{1}^{2}}{\kappa_{1}^{2}}\dfrac{\omega^{2}}{c^{2}}  \dfrac{d}{\sinh^{2} \kappa_{1}d}\right. \\ \qquad \left.+k^{2}\dfrac{\varepsilon_{1} }{\kappa_{1}^{3}}\coth \kappa_{1}d + k^{2} \dfrac{\varepsilon_{2}}{\kappa_{2}^{3}} \right] \;,\end{multline}
\begin{multline} \label{25}
\left\langle U_{\mathrm{sp}}\right\rangle =\dfrac{\varepsilon_{0}}{4}A_{-}^{2}\sinh^{2} \kappa_{1}d\frac{\pi}{\varepsilon_{0}D_{\mathrm{g}}} \frac{\omega^{2}+v_{\mathrm{s}}^{2}k^{2}}{\omega^{2}}\vert\sigma_{\mathrm{g}}\vert^{2}  \;.\end{multline}
Thus, for the surface plasmon strength function $\Theta_{\mathrm{sp}}$ and the electromagnetic strength function
$\Theta_{\mathrm{ph}}$ we obtain 
\begin{multline} \label{26} \Theta_{\mathrm{ph}}=\dfrac{\left\langle U_{\mathrm{ph}}\right\rangle}{\left\langle U\right\rangle}\\=\dfrac{\frac{\varepsilon_{1}^{2}}{\kappa_{1}^{2}}\frac{\omega^{2}}{c^{2}}  \frac{d}{\sinh^{2} \kappa_{1}d}+k^{2}\frac{\varepsilon_{1} }{\kappa_{1}^{3}}\coth \kappa_{1}d + k^{2} \frac{\varepsilon_{2}}{\kappa_{2}^{3}}  }{ \frac{d \frac{\varepsilon_{1}^{2}}{\kappa_{1}^{2}}\frac{\omega^{2}}{c^{2}} }{\sinh^{2} \kappa_{1}d}+k^{2}\frac{\varepsilon_{1} }{\kappa_{1}^{3}}\coth \kappa_{1}d + k^{2} \frac{\varepsilon_{2}}{\kappa_{2}^{3}}+\frac{\pi \vert\sigma_{\mathrm{g}}\vert^{2}}{\varepsilon_{0}D_{\mathrm{g}}} \frac{\omega^{2}+v_{\mathrm{s}}^{2}k^{2}}{\omega^{2}}  }\;. \end{multline}
\begin{multline} \label{27} \Theta_{\mathrm{sp}}=\dfrac{\left\langle U_{\mathrm{sp}}\right\rangle}{\left\langle U\right\rangle}\\=\dfrac{\frac{\pi}{\varepsilon_{0}D_{\mathrm{g}}} \frac{\omega^{2}+v_{\mathrm{s}}^{2}k^{2}}{\omega^{2}}\vert\sigma_{\mathrm{g}}\vert^{2}  }{ \frac{d \frac{\varepsilon_{1}^{2}}{\kappa_{1}^{2}}\frac{\omega^{2}}{c^{2}} }{\sinh^{2} \kappa_{1}d}+k^{2}\frac{\varepsilon_{1} }{\kappa_{1}^{3}}\coth \kappa_{1}d + k^{2} \frac{\varepsilon_{2}}{\kappa_{2}^{3}}+\frac{\pi \vert\sigma_{\mathrm{g}}\vert^{2}}{\varepsilon_{0}D_{\mathrm{g}}} \frac{\omega^{2}+v_{\mathrm{s}}^{2}k^{2}}{\omega^{2}}  }\;. \end{multline}
Fig. \ref{fig.5} shows the variation of $\Theta_{\mathrm{ph}}$ and $\Theta_{\mathrm{sp}}$ with respect to the dimensionless wavenumber of the screened plasmon-polaritons of graphene near a PEC. One can see that for low values of $D$, a plasmonic wave of the system is largely plasmon-like and the role of the surface electromagnetic wave is small compared with that of the surface plasmon wave. Only for high values of $D$ the strength functions of surface plasmon and surface electromagnetic wave are comparable. 

\section{Conclusion} 
In summary, we have investigated the properties of screened plasmons of graphene near a PEC based on classical electrodynamics and the linearized hydrodynamic model with Fermi correction. In this way, at first, we have derived the dispersion relation of the mentioned screened plasmonic waves. We have studied numerically the effects of graphene distance from the PEC and graphene sheet carrier density on the surface wave properties of the system. Numerical results for the present system indicate alterations in the physical behavior of the surface waves, in comparison with those obtained for an isolated monolayer graphene. Also, we have derived the analytical expressions for the power flow, energy density, and energy velocity of screened plasmons of the system. Furthermore, we have obtained the analytical expressions for the damping function, and surface plasmon and electromagnetic field strength functions of surface waves of the system with small intrinsic damping. The results show that the plasmonic properties of graphene near the PEC may present new behaviors that make it suitable for applications in the mid-infrared to a few THz range of frequencies.

\section*{ACKNOWLEDGMENTS}
The first-named author would like to thank the Department of Electronic and Communications Engineering at the Yildiz Technical University for its hospitality during his visit. Also, A.M. would like to acknowledge the financial support of the Kermanshah University of Technology for this research opportunity under grant number S/P/F/6.

\section*{AUTHOR DECLARATIONS}
\subsection*{Conflict of Interest}

The authors have no conflicts to disclose.

\subsection*{Author Contributions}

\textbf{Afshin Moradi:} Project administration (equal); Conceptualization (lead); Investigation (lead);
Methodology (lead); Validation (lead); Formal analysis (lead); Software (lead); Writing - original draft
(lead); Writing - review and editing (equal). \textbf{Nurhan T\"urker Tokan:} Project administration (equal); Conceptualization (supporting); Investigation (supporting);
Methodology (supporting); Validation (supporting); Formal analysis (supporting); Writing - review and editing (equal).

\subsection*{DATA AVAILABILITY} 

The data that supports the findings of this study are available within the article.

\renewcommand{\theequation}{A-\arabic{equation}} 
\setcounter{equation}{0}
\section*{Appendix: The energy density and power flow on graphene with Fermi correction}
From the Poynting theorem for energy in standard electrodynamics, we have: \b \label{A1} \nabla\cdot
\textbf{S}+\partial_{t}u=-\textbf{E}\cdot\textbf{J}\;, \e where
$\textbf{S}=\textbf{E}\times\textbf{H}$ is known as the Poynting
vector, which is a power density vector associated with an
electromagnetic field and
$u=\frac{1}{2}\left(\textbf{E}\cdot\textbf{D}+\textbf{H}\cdot\textbf{B}\right)$
is the energy density of electromagnetic waves. For the present
nonmagnetic system, we have $\textbf{D}=\varepsilon_{0}\varepsilon\textbf{E}$
and $\textbf{B}=\mu_{0}\textbf{H}$ (where $\mu_{0}$ is the
permeability of free space). At this
stage, by employing Eqs. \eqref{1} and \eqref{2} and after doing some algebra,
we rewrite the right-hand side of Eq. \eqref{A1} on graphene surface as:
\b \label{A2}
\textbf{E}\cdot\textbf{J}=\frac{\pi }{2D_{\mathrm{g}}}\partial_{t}j_{x}^{2}+\frac{\pi e^{2}}{2D_{\mathrm{g}}}v_{\mathrm{s}}^{2}
\partial_{t}n_{\mathrm{g}}^{2}-\frac{\pi e}{D_{\mathrm{g}}} v_{\mathrm{s}}^{2}\partial_{x}\textbf{e}_{x}\cdot
n_{\mathrm{g}}j_{x}\textbf{e}_{x}\;.\e
Then,
this equation together with Eq. \eqref{A1} yields the energy density
$U_{\mathrm{g}}$ and the power flow $S_{\mathrm{g}}$ on the graphene
surface in the forms, as: \b \label{A3} U_{\mathrm{g}}=U_{\mathrm{gk}}+U_{\mathrm{gp}}\;,\e \b \label{A4}
S_{\mathrm{g}}=-\frac{\pi e}{D_{\mathrm{g}}} v_{\mathrm{s}}^{2} n_{\mathrm{g}}j_{x}\;.\e
On the right-hand side
of Eq. \eqref{A3}, the first term is the kinetic-energy density
$U_{\mathrm{gk}}=\pi j_{x}^{2}/2D_{\mathrm{g}}$ and the second term represents the
the potential-energy density $U_{\mathrm{gp}}=\pi e^{2} v_{\mathrm{s}}^{2} n_{\mathrm{g}}^{2}/2D_{\mathrm{g}}$.

\end{document}